\documentclass[twocolumn,twoside]{article}

\usepackage{a4}
\usepackage{amssymb}
\usepackage{amsmath}
\usepackage[numbers,sort&compress]{natbib}
\usepackage{graphicx}
\usepackage{blindtext}

\begin{document}

\title{\bf {\Large  First Adiabatic Invariant and the Brightness Temperature of Relativistic Jets}} 

\date{{\normalsize\textit{$^{1}$P N Lebedev Physical Institute, Russian Academy of Sciences,
Leninskii prosp. 53, 119991, Moscow, Russia,\\
$^{2}$Moscow Institute of Physics and Technology
(National Research University), Institutskii per. 9,
141701 Dolgoprudny, Moscow region, Russia }} \\[1ex]
{\small \textit{Pis’ma v Astronomicheskii Zhurnal}, 2023, Vol. 49, No. 3, pp. 197–207. [in Russian]\\
English translation: \textit{ISSN 1063-7737, Astronomy Letters} 
2023, Vol. 49, No. 3, pp. 193–202.  Pleiades Publishing, Inc., 2023.  Translated by V. Astakhov}
\\.
\\
{\bf Abstract.} Assuming that the first adiabatic invariant for radiating particles in relativistic jets is conserved, we investigate the change in brightness temperature along the jet axis. We show that in this case the observed break in the dependence of the brightness temperature on the distance to the ”central engine” can be explained.}

\author{V. S. ~Beskin$^{1,2}$, T. I. Khalilov$^{1,2}$, and V. I. ~Pariev$^{1,2}$}

\maketitle

\setcounter{secnumdepth}{3}
\setcounter{tocdepth}{2}

DOI: 10.1134/S1063773723030015 

\section*{Introduction}

Relativistic jets from active galactic nuclei (AGNs)
are apparent manifestations of their activity at an
early stage of evolution (Begelman et al. 1984; Urry
and Padovani 1995; Davis and Tchekhovskoy 2020;
Komissarov and Porth 2021). Their hydrodynamic
velocities correspond to Lorentz factors
$\Gamma \sim$ 10. In the galaxy M87
this motion is observed directly, while the Lorentz
factor is $\Gamma \approx 6$ (Junor and
Biretta 1995). In many cases, the outflowing plasma
retains relativistic velocities at enormous distances
from the nucleus before decelerating noticeably when
interacting with the intergalactic medium. There
is no doubt that the observed radio emission of the
jet is associated with the synchrotron radiation of
relativistic electrons. This is suggested both by
the power-law spectrum of the observed emission
and by its characteristic low-frequency cutoff easily
explainable by synchrotron self-absorption (Lyutikov
et al. 2003, 2005; Porth et al. 2011; Sokolovsky
et al. 2011).

At present, owing to the development of very long
baseline interferometry (VLBI), it is possible to trace
the properties of the radio emission up to the innermost (of the order of hundreds of gravitational radii)
jet regions (Kovalev et al. 2005). In particularly, it
has been shown recently that in many cases the jet
shape changes from parabolic to conical (Asada and
Nakamura 2012; Kovalev et al. 2020; Park et al. 2021;
Boccardi et al. 2021), so that this phenomenon can be said to be universal for relativistic jets. However,
despite the large volume of accumulated information,
many issues still await their resolution. In particular, this concerns the recently detected break in the
dependence of the brightness temperature $T_{\rm br}$ on the
distance to the ”central engine” $z$ (Kadler et al. 2004;
Baczko et al. 2019; Burd et al. 2022) that occurs
in many cases at distances \mbox{$\sim$ 1 pc,} i.e., exactly in
the region of the transition from parabolic to conical
shape. This break is observed already in dozens of
objects, with the exponents $a$ in the power law $T_{\rm br} \propto z^{-a}$ being confined in a wide range,
\begin{eqnarray}
a_{1} & = & 3.0 \pm 1.0, 
\label{Icon1}\\
a_{2} & = & 2.7 \pm 1.0,
\label{Icon2}
\end{eqnarray}
where $a_{1}$ and $a_{2}$ correspond to small and large distances from the central engine, respectively. Such a
wide spread of parameters, when close mean values
contain little information, implies that in some cases
$a_{1} > a_{2}$ (upward break), while in other cases $a_{1} < a_{2}$
(downward break). At the same time, no difference
in the exponents a for quasars and BL Lac objects is
observed.

One of the reasons restraining the construction
of a consistent theory of radio emission from jets
was that the energy of radiating particles should be
much greater than the energy of hydrodynamic motion. Therefore, for many years it had been impossible to directly connect the questions related to the
observed radio emission with the magnetohydrodynamic (MHD) theory of jets that had already been
developed for several decades (Heyvaerts and Norman 1989; Pelletier and Pudritz 1992; Beskin 2006; Tchekhovskoy et al. 2008; McKinney et al. 2012),
since the MHD theory of jets told us nothing about
the energetics of radiating particles. Since so far
there had been no consensus on the formation of the
spectrum of radiating particles (see, e.g., Marscher
and Gear 1985; Istomin and Pariev 1996; Pariev
et al. 2003), the uncertainty arising in this link did not
allow the evolution of the emission parameters along
the jet axis to be investigated self-consistently.

As a matter of fact, only the adiabatic model of Marscher (1980) 
is currently a sufficiently well-developed model that allows the 
dependence of the spectrum of radiating particles on the distance to
the central engine $z$ to be analyzed. Indeed, using the relativistic 
equation of state $PV^{4/3} =$ const for a conical jet, when the size 
of the emitting regions $R \propto z$, we immediately obtain 
$\gamma \propto R^{-2/3}$ for the average Lorentz factor. These 
approximations were generalized to the situation of accelerating 
jets already in Lobanov and Zensus (1999). Subsequently, based
on the results presented above, Lobanov et al. (2000) derived 
a dependence of the brightness temperature on the transverse size
of the jet in the form $T_{\rm br} \sim R^{-\xi}$ ($\xi \approx 2$), 
where $R$ is the transverse size of the jet. For example, such a 
dependence was applied for the observed changes in brightness 
temperature in Gom{\' e}z et al. (2016) and Nair et al. (2019).

However, it is not obvious that such a simple hydrodynamic model 
will be valid for a strongly magnetized flow as well. In any case, 
for radiating particles their mean free path $l_{\gamma}$ (Berestetskii et al. 1989),
\begin{equation}
l_{\gamma} \sim \frac{\gamma^2}{r_{\rm e}^2n_{\rm }} \sim 10^{8} \, R_{\rm L} 
\left(\frac{\lambda}{10^{12}}\right)^{-1}
\left(\frac{\gamma}{10^3}\right)^2
\left(\frac{B_{\rm p}}{10^{}\rm{G}}\right)^{-1},
\end{equation}
turns out to be greater than the characteristic size of
the system $L$. Here, $\lambda$ is a convenient dimensionless
coefficient (production multiplicity) that parameterizes the particle number density
\begin{equation}
n_{\rm e} = \lambda n_{\rm GJ}\label{nGJ},
\end{equation}
where $n_{\rm GJ} = \Omega_{\rm F}B_{\rm p}/(2\pi \, c \, e)$ is so-called Goldreich-Julian number density, 
i.e., the minimum particle number density needed for the approximation of ideal magnetohydrodynamics to hold, 
and $\Omega_{\rm F} \approx \Omega_{\rm H}/2$, where $\Omega_{\rm H}$ is the black hole rotation rate. 
Finally, $R_{\rm L} = c/\Omega_{\rm F}$ is the radius of the light cylinder, which exceeds the
black hole radius approximately by a factor of 10
(Davis and Tchekhovskoy 2021; Komissarov and
Porth 2021). We used a poloidal magnetic field
$B_{\rm p} = 10^2$ G typical for the scale of the light cylinder.
As regards $\lambda$, below we set $\lambda \sim 10^{12}$. Such a
large value follows both from numerical simulations of particle production in the magnetospheres of black
holes (Mo{\'s}cibrodzka et al. 2011) and from observations (Nokhrina et al. 2015).

Of course, the condition $l_{\gamma} \gg L$ is insufficient for
the relation $PV^{4/3} =$ const to break down; this requires that the distribution function be isotropized
sufficiently slowly, while this process depends on the
level of outflowing plasma turbulence, about which
there is currently no reliable information. Therefore,
below we consider a two-component model consisting of a hydrodynamic flow with a small spread
of particles in energies (for which the isotropization condition is fulfilled) and a high-energy tail of
weakly interacting particles with which the observed
synchrotron radiation is associated. To explain the
observed radiation (including the optical and X-ray
one), it is commonly assumed that the power-law
spectrum can extend to energies at least of the order
of several TeV. The number density of radiating particles should be noticeably lower than the background
plasma number density.

It is natural to assume that the first adiabatic invariant $I_{\perp} = p_{\perp}^2/B$ will be conserved for radiating particles. In this case, it will also be possible
to relate their energetics to the jet parameters, which,
as has already been noted, are satisfactorily modeled
within present-day MHD models. As a result, a
unique opportunity to obtain direct information
about the evolution of the emitting plasma properties
along the jet axis appears. In particular, it becomes
possible to check whether an additional particle acceleration within the jet is necessary to explain the
intensity of the observed radiation.

In this paper, based on our analysis of the motion
of charged particles in conical and parabolic relativistic jets, we investigate the change in brightness
temperature along the jet axis. We show that in
this case the observed break in the dependence of
the brightness temperature on the distance to the
central engine can be explained. Good agreement
with observations is achieved without any additional
particle acceleration within the jet itself.

In the first part we discuss two models of relativistic jets 
that describe quite adequately the internal structure of 
relativistic jets from AGNs. The second part is devoted to 
analyzing the conservation of the first adiabatic invariant 
in the crossed electric and magnetic fields of relativistic 
jets. We show that the first adiabatic invariant is
conserved with a great accuracy. Finally, in the third
part we analyze the pattern of change in brightness
temperature along the jet axis through a different
dependence of the behavior of the radiating particle
spectrum due to the conservation of the first
adiabatic invariant. We show that this effect can serve
as a basis for explaining the observational data.

\section*{Two models of relativistic jets}

As has already been noted above, the observations
themselves suggest that conical and parabolic flows
may be chosen as a fairly good geometrical model
of relativistic jets. Therefore, as a basis, below we consider two simple analytical models of forcefree relativistic jets --- the conical (to be more precise,
quasi-spherical) solution of Michel (1973) and the
parabolic solution of Blandford (1976). Here an important help to us is the fact that, according to numerical simulations (see, e.g., McKinney et al. 2012), a
substantial region near the jet axis does have a regular
magnetic field.

The electromagnetic fields of Michel’s conical
force-free flow in spherical coordinates $r$, $\theta$, $\varphi$ are
\begin{eqnarray}
B_{r} & = & B_{\rm L}\frac{R_{\rm L}^2}{r^2}, 
\label{Conical_field_1} \\ 
B_{\varphi} & = & -B_{\rm L}\frac{R_{\rm L}}{r} \sin\theta, 
\label{Conical_field_2}\\
E_{\theta} & = & - (1 - \varepsilon) B_{\rm L}\frac{R_{\rm L}}{r} \sin\theta, 
\label{Conical_field_3}
\end{eqnarray}
where again $R_{\rm L}$ is the radius of the light cylinder and
$B_{\rm L}$ is the magnetic field at $r = R_{\rm L}$. We assume
that this solution exists only in the narrow region
$\theta < \theta_{\rm jet} \sim 0.1$ near the jet axis. In addition, the factor
$\left(1 - \varepsilon\right)$ was added to the expression for the electric
field, where the constant $\varepsilon \ll 1$ allows one to simulate
the absence of particle acceleration at great distances,
when the entire electromagnetic energy flux has been
transferred to the plasma flow. In other words, the
small parameter $\varepsilon$ is responsible for the saturation of
the particle energy (Lorentz factor).

Indeed, using the fundamental theoretical result
that in an asymptotically distant region beyond the
light cylinder the particle energy in the hydrodynamic
component{\footnote{For the high-energy component this is not the case (see
Prokofev et al. 2015).}} approaches the energy corresponding to
drift motion (Tchekhovskoy et al. 2008; Beskin 2010;
Bogovalov 2014)
\begin{equation}
{\bf U}_{\rm dr} = c \frac{{\bf E}\times{\bf B}}{B^2},   
\end{equation}
and, therefore, the longitudinal velocity along the
magnetic field may be neglected when determining
the hydrodynamic velocity of the particles, for the
hydrodynamic Lorentz factor we obtain
\begin{eqnarray}
\Gamma = \frac{1}{\sqrt{1 - v^2/c^2}} =
\left(1 - \frac{E^2}{B_{\varphi}^2 + B_{\rm p}^2}\right)^{-1/2} \approx \nonumber\\
\approx \left(1 - \frac{1 - 2\varepsilon}{1 + 1/x^2}\right)^{-1/2} \approx
\left(2\varepsilon + \frac{1}{x^2}\right)^{-1/2}.
\label{Gamma}
\end{eqnarray}
Here, $x = r\sin\theta/R_{\rm L}$ is the dimensionless distance to
the jet axis. As a result, at small distances from the
central engine, i.e., at $x < (2\varepsilon)^{-1/2}$, we obtain
\begin{equation} 
\Gamma \approx x,
\label{Gx}
\end{equation}
i.e., the well-known asymptotic behavior for collimated magnetized jets. On the other hand, at great distances,
i.e., at $x > (2\varepsilon)^{-1/2}$, we have $\Gamma \approx (2\varepsilon)^{-1/2} \approx$ const.
This asymptotic solution simulates the saturation region,
when the entire energy flux is concentrated in the
hydrodynamic particle flow. Therefore, we can write
\begin{equation} 
\varepsilon \approx \frac{1}{2\sigma_{\rm M}^2},
\label{vareps}
\end{equation}
where $\sigma_{\rm M}$ is the so-called Michel (1969) magnetization parameter that has the meaning of a maximally possible Lorentz factor. According to Nokhrina et al. (2015), for most of the relativistic jets from
AGNs \mbox{$\sigma_{\rm M} = 10$--$50$}, consistent with the values determined from superluminal motions. Thus, the structure of the electromagnetic fields unambiguously determines all of the hydrodynamic flow characteristics needed for us below.

Another model is the force-free solution with
a parabolic poloidal magnetic field found by Blandford (1976). In spherical coordinates $r$, $\theta$, $\varphi$ it can be written as (Beskin 2006)
\begin{equation} 
{\bf B}_{\rm p} = B_{\rm L}\frac{\nabla X \times {\bf e}_{\varphi}}{\sin\theta\sqrt{1+\Omega_{\rm F}(X)^2 X^2/c^2}}\frac{R_{\rm L}}{r}\label{parabolBpol}\mbox{,}
\end{equation} 
where $R_{\rm L} = c/\Omega_{\rm F}(0)$ and
\begin{equation} 
X = r(1 - \cos\theta),
\end{equation} 
so $|\nabla X| = \sqrt{2 - 2\cos\theta}$. The condition \mbox{$X(r, \theta) = $ const} ($\theta \propto r^{-1/2}$) corresponds to a parabolic structure whereby all field lines pass through the equatorial
plane (with the central part of the field lines crossing
the black hole horizon). The radius of the light
cylinder $R_L$ and the value of the function $\Omega_F$ at
$X=0$ are related by the condition $R_{\rm L} = c/\Omega_{\rm F}(0)$.
Accordingly, the electric and toroidal magnetic fields are
\begin{eqnarray}
{\bf E} = -\frac{B_{\rm L}(1 - \varepsilon)}{\sqrt{1 + \Omega_{\rm F}(X)^2 X^2/c^2}} \,  \times \nonumber \\
\times \left(\frac{\Omega_{\rm F}(X)R_{\rm L}}{c}\right) \, \nabla X, 
\label{parabolE} \\
B_{\varphi} = -\frac{B_{\rm L}}{\sqrt{1 + \Omega_{\rm F}(X)^2 X^2/c^2}} \times \nonumber \\
\times \frac{(2 - 2\cos\theta)}{\sin\theta}\, \left(\frac{\Omega_{\rm F}(X)R_{\rm L}}{c}\right).
\label{parabolB}
\end{eqnarray}
Here, $\Omega_{\rm F}$, the so-called angular velocity of the magnetic field lines, must depend on $X$, since only those
configurations for which $\Omega\left(X\right) X/c < 1$ turn out to
be possible (otherwise the central engine would rotate with a speed exceeding the speed of light); in
all of the subsequent calculations we assumed that
$\Omega\left(X\right) X/c = 0.5$ at $X > 0.5 \, R_{\rm L}$. Here, by analogy
with the previous model, the constant factor $\left(1 - \varepsilon\right)$
was added to the electric field. It is easy to verify that
for this configuration the asymptotic behavior $\Gamma = x$ (\ref{Gx}) also
holds in the region of a strongly magnetized flow and $\Gamma \approx$ const in the saturation region.

\section*{Conversation of the first adiabatic invariant}
Before turning to our main problem, i.e., finding
the brightness temperature of relativistic jets, let us
discuss in detail the properties of the motion of individual particles 
in the electromagnetic fields defined above. First of all, it is necessary to check
whether the first adiabatic invariant
\begin{equation}
I_{\perp} = \frac{(p_{\perp}^{\prime})^2}{h}.
\label{Iperp}
\end{equation}
is indeed be conserved during such a motion. Here,
\begin{equation}
h = \sqrt{B^2 - E^2}
\end{equation}
is the magnetic field in the hydrodynamic rest frame
(the electric field is zero). Accordingly, $p_{\perp}^{\prime}$ is the
transverse momentum also in the rest frame. The
point is that in the standard formulation the motion
of a particle is considered in a stationary magnetic
field, i.e., in a specified inertial reference frame. In
contrast, in our case the comoving reference frame
(the reference frame in which the plasma is at rest)
accelerates in inhomogeneous crossed electric and magnetic fields.
Therefore, it is useful to check the conservation of
the first adiabatic invariant in such a nontrivial case.

\begin{figure*}[!ht]
	\begin{minipage}{0.5\linewidth}
		\center{\includegraphics[width=1\linewidth]{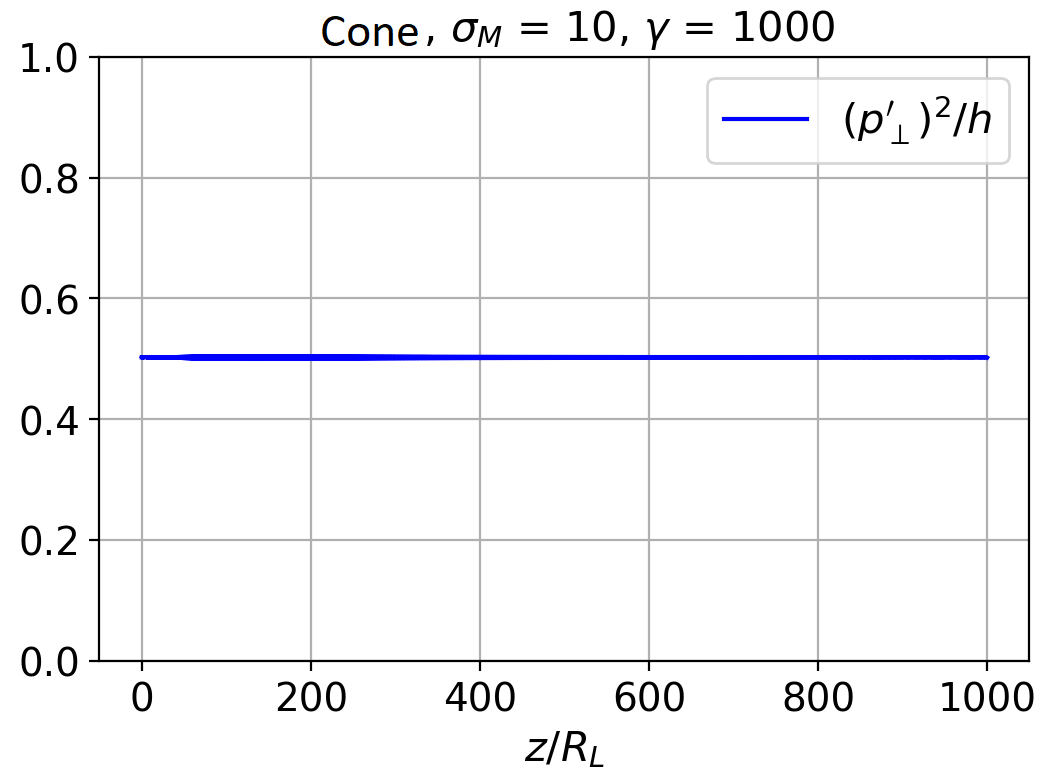} }
	\end{minipage}
	\hfill
	\begin{minipage}{0.5\linewidth}
		\center{\includegraphics[width=1\linewidth]{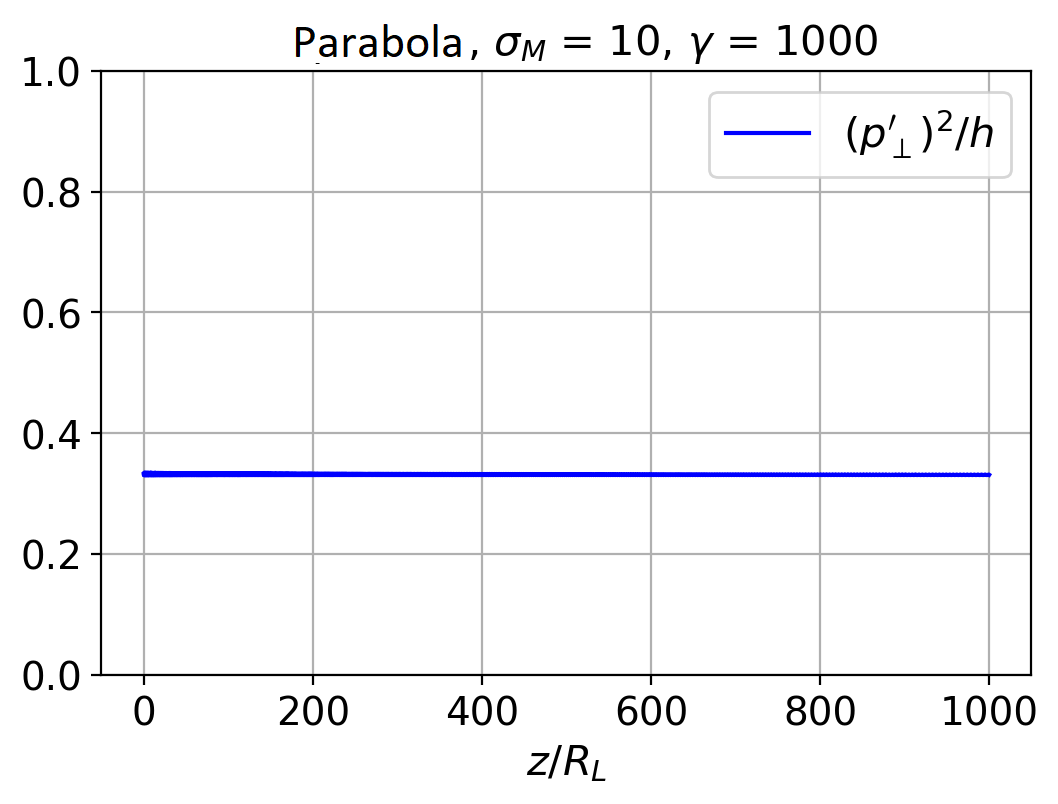} }
	\end{minipage}
 	\caption{. Conservation of the first adiabatic invariant in the comoving reference frame (relative units) for conical and
parabolic flows.}
\label{fg1}	
\end{figure*}

Figure \ref{fg1} shows the results of our numerical integration of the motion of 
particles in the electromagnetic fields (\ref{Conical_field_1})--(\ref{Conical_field_3}) 
and (\ref{parabolE})--(\ref{parabolB}) for typical parameters of relativistic jets, 
$\sigma_{\rm M} = 10$ and the Lorentz factor $\gamma = 10^{3}$; the value of the small 
parameter $\varepsilon$ is determined from relation (\ref{vareps}). As we see, the first
adiabatic invariant $I_{\perp}$ (\ref{Iperp}) is indeed conserved with a good accuracy
for both conical and parabolic geometries. Since for ultrarelativistic radiating particles 
\mbox{$m_{\rm e}c\gamma^{\prime} \approx p_{\perp}^{\prime}$,} we find that
\begin{equation}
\gamma^{\prime} = \frac{I_{\perp}^{1/2}}{m_{\rm e}c} \, h^{1/2}.
\label{gamma-h}
\end{equation}

Here, it is also important to draw attention to the fact that the invariant 
$I_{\perp}$ does not change while the pattern of the dependence of $h$ on the 
distance $z$ to the central engine changes significantly. Indeed, it is easy 
to verify, using relation (\ref{Gamma}), that for a conical flow at small distances, 
i.e., at $x < 1/\sqrt{2\varepsilon}$, the asymptotic behavior $h \propto z^{-2}$ 
is valid, while at great distances $x > 1/\sqrt{2\varepsilon}$ the asymptotic 
behavior $h \propto z^{-1}$ is valid. Accordingly, for a parabolic flow at small
distances, i.e., at $x < 1/\sqrt{2\varepsilon}$, we have $h \propto z^{-1}$ and
$h \propto z^{-1/2}$ at great distances $x > 1/\sqrt{2\varepsilon}$. As we show, 
it is the presence of such a break that allows us to explain the pattern of change 
in the dependence of the brightness temperature on the distance $z$ to the
central engine.

\begin{figure*}[!ht]
	\begin{minipage}{0.5\linewidth}
		\center{\includegraphics[width=1\linewidth]{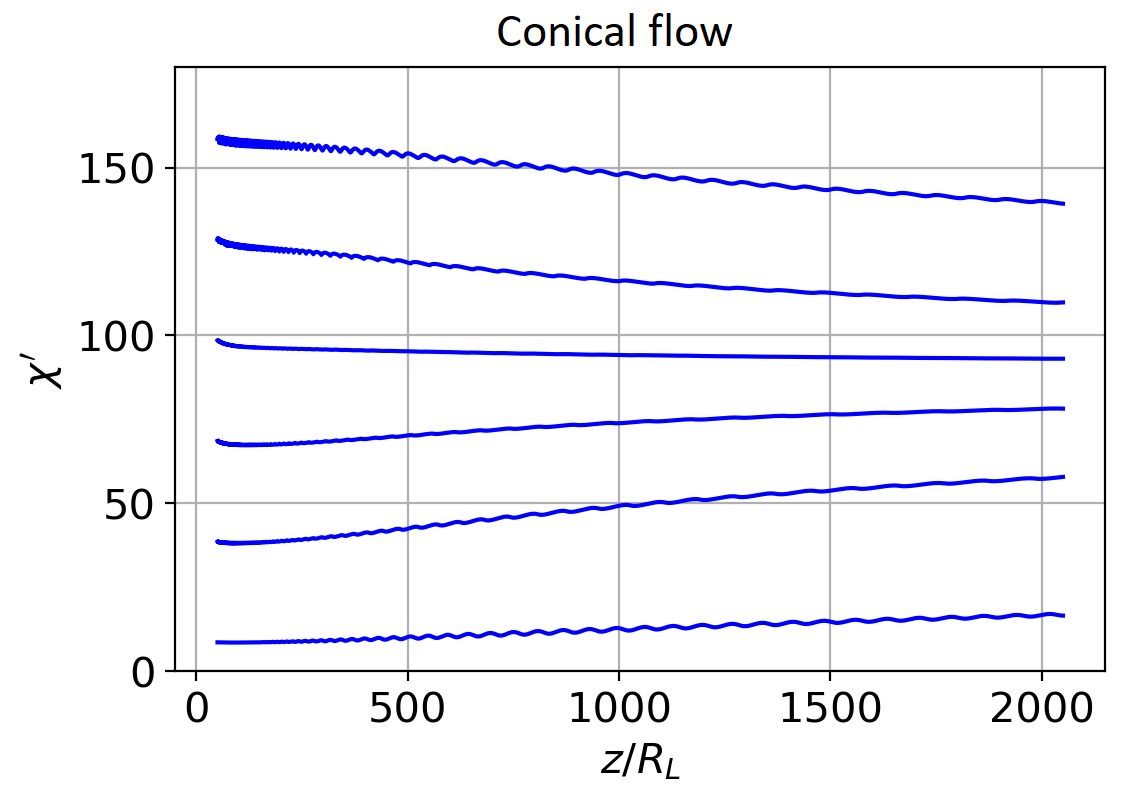} }
	\end{minipage}
	\hfill
	\begin{minipage}{0.5\linewidth}
		\center{\includegraphics[width=1\linewidth]{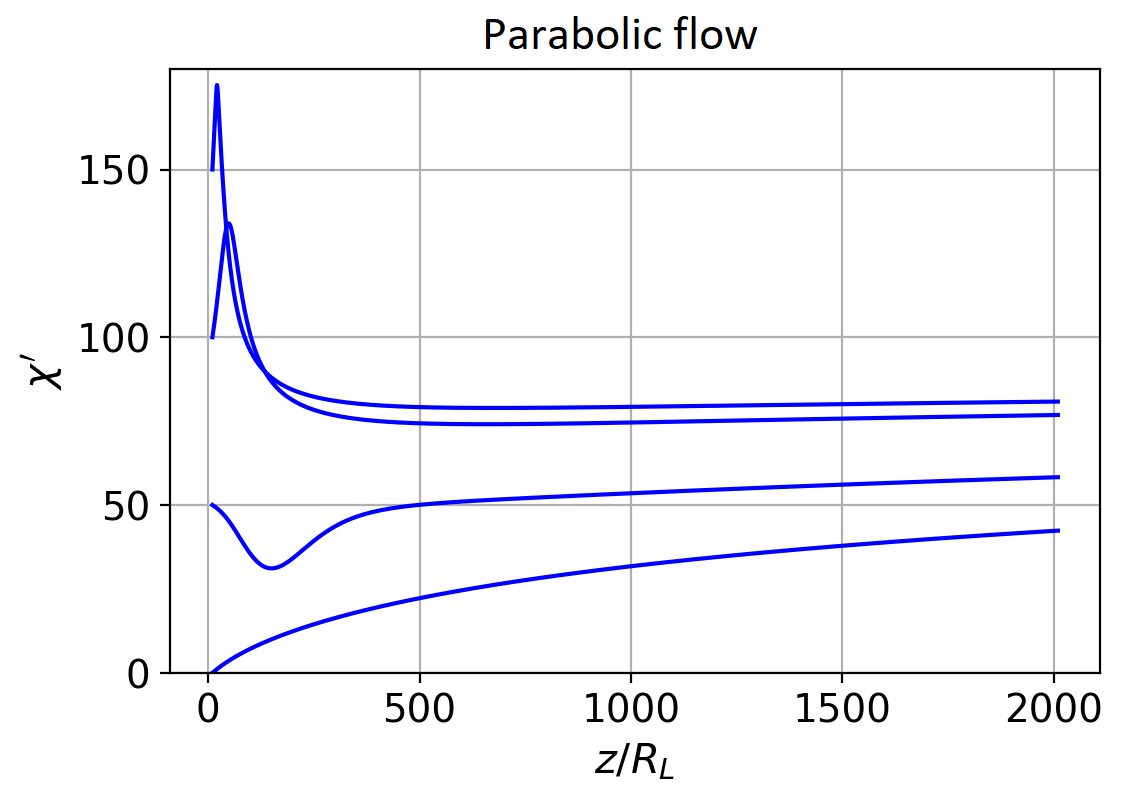} }
	\end{minipage}
 	\caption{Pitch angle of an individual particle $\chi^{\prime}$ in the comoving reference frame versus distance $z$ to the central engine.}
\label{fg2}	
\end{figure*}

Finally, note yet another important circumstance
that we need below. As shown in Fig. \ref{fg2}, in the
comoving reference frame the pitch angle of an individual particle $\chi^{\prime}$ does not decrease with decreasing
magnetic field (i.e., with increasing distance $z$ to the
central engine), as is sometimes the case in static,
purely magnetic configurations, but, on the contrary,
though slowly, tends to $90^{\circ}$. This once again confirms
the previously noted result that in magnetized winds
(and in the comoving reference frame) the longitudinal (parallel to the magnetic field) velocity component
may be neglected compared to the drift velocity. It is
clear that this effect takes place only beyond the light
cylinder, where the role of the electric field becomes
decisive. In contrast, within the light cylinder, as in
the absence of an electric field, the conservation of the
first adiabatic invariant leads to a decrease in
the pitch angle $\chi^{\prime}$
. Such a decrease in $\chi^{\prime}$ is clearly
seen in Fig. \ref{fg2} for a parabolic field at the initial phase
of the particle trajectories, as long as they are in the
immediate vicinity of the light cylinder. As has already
been noted, the various instabilities inherent in highly
anisotropic distributions tend to reduce the degree of anisotropy. Therefore, the assumption about
slow isotropizatrion is a necessary condition for the
model being considered here. Only the fact that the
pitch angles do not tend to $0^{\circ}$, when the synchrotron
radiation power is substantially suppressed, will be
important to us below.

\section*{Brightness temperature}

Let us now look at how the change in the energy
of radiating particles that inevitably arises as the jet
expands due to the conservation of the first
adiabatic invariant affects the change in brightness
temperature $T_{\rm br}$ along the jet axis. Here we use the
standard relations derived in Lyutikov et al. (2003,
2005). A significant difference lie only in the fact
that we take into account both the jet expansion
and the evolution of the spectrum of radiating particles. At the same time, we ignore the synchrotron losses in the high-energy part of the spectrum (see
the Appendix).

Note at once that the possibility to restrict ourselves to the approximation of an optically thin plasma,
in which neither the synchrotron self-absorption nor
the Faraday rotation is taken into account, stems
from the fact that the saturation region of interest
to us here is sufficiently far from the central engine,
where the plasma density and the magnetic fields
are already not so great. As is well known, in
the centimeter wavelength range there is noticeable
self-absorption determined from the core shift (see,
e.g., Nokhrina et al. 2015) at distances no greater
than several hundred gravitational radii. Regarding
the Faraday rotation, as is well known, even the
rotation measures ${\rm RM} \sim 100$ rad m$^{-2}$ recorded in the
innermost regions (Kravchenko et al. 2017; Gabuzda
et al. 2017) do not lead to noticeable depolarization in
the centimeter wavelength range.

Below we will present the corresponding calculations, while here we will begin our discussion with
the expression for the number density of radiating
particles in the comoving reference frame $n_{\gamma}^{\prime}$ with a power-law energy spectrum in the range $\gamma_{0} < \gamma^{\prime} < \gamma_{\rm max}$:
\begin{equation}
{\rm d}n_{\gamma}^{\prime} = K_{\rm e} \, (\gamma^{\prime})^{-p}  \, {\rm d}\gamma^{\prime} \, {\rm d}^3{r}^{\prime} \, {\rm d}\Omega.
\label{ne}
\end{equation}
Here, $p$ is the spectral index, ${\rm d}\Omega$ is an element of
the solid angle, and $K_{\rm e}$ is the normalization constant.
Note at once that in our model it is important that the
lower limit integration plays a key role. As we show below,
the lower limit can also correspond to nonrelativistic
velocities ($\gamma_{0} = 1$); our main conclusions will not
change because of that.

Next, note that owing to the tendency for the pitch
angle $\chi^{\prime}$ to increase, which is pointed out above, in this paper we will ignore the dependence of the distribution
function (\ref{ne}) on the solid angle ${\rm d}\Omega$. This is because
at small angles between the jet axis and the direction to the observer typical for quasars and owing to the
nearly toroidal magnetic field in the main part of the
jets, the synchrotron radiation beam of most radiating
particles will be oriented toward the observer.

As a result, since, as has been shown above, the
energy of all radiating particles when propagating
along the expanding jet changes proportionally to $h$,
the power-law shape of the spectrum is retained
along the entire jet. On the other hand, if the normalization of the particle energy spectrum is chosen in the form $\int f(\gamma^{\prime}) {\rm d}\gamma^{\prime} = 1$, then in this case
\begin{equation}
K_{\rm e} = (p - 1) n_{\gamma}^{\prime}({\bf r}^{\prime}) \, \gamma_{0}^{p - 1}({\bf r}^{\prime}).\label{Ke_exp}
\end{equation}
Here, we assumed that $\gamma_{\rm max}^{\prime} \gg \gamma_{0}$ and $\gamma_{0} \gg 1$. Owing to the main relation $\gamma_{0} \propto h^{1/2}$ (\ref{gamma-h}), the normalization factor $K_{\rm e}$ acquires a dependence on the
invariant $h$. This allowance for the dependence of $K_{\rm e}$
on the coordinates in agreement with the conserved
adiabatic invariant $I_{\perp}$,
\begin{equation}
K_{\rm e} \propto n_{\gamma}^{\prime} \,  h^{(p-1)/2},
\label{Ke}
\end{equation}
due to the dependence of the magnetic field in the
comoving reference frame $h$ on distance $z$, that is
the subject of our study. It is easy to verify that the
dependence (\ref{Ke}) will also hold if the lower boundary
of the spectrum of radiating particles is close to $m_{\rm e}c^2$ ($\gamma_{0} = 1$). In this case, the change in $K_{\rm e}$ is related
to the change in the number of radiating particles with
$\gamma^{\prime} > 1$.

Using now the standard expression for the intensity $I_{\nu}$ (Lyutikov et al. 2005), we obtain
\begin{eqnarray}
{\rm d}I_{\nu} = 2\pi\frac{(p + 7/3)}{(p + 1)} \,\kappa(\nu) \, \frac{{\rm d}S \,
{\rm d}l}{D^2} \, \times \nonumber \\
\times {\cal D}^{2 + (p -1)/2} \,  |h\sin{\hat \chi}|^{(p + 1)/2} \,  {\rm
d}\nu.
\end{eqnarray}
Here, $\hat\chi$ is the angle between the magnetic field and
the line of sight,
\begin{eqnarray}
\kappa(\nu) = \frac{\sqrt{3}}{4} \Gamma\left(\frac{3p - 1}
{12}\right)\Gamma\left(\frac{3p + 7}{12}\right)\, \times \nonumber\\\
\times \frac{e^3}{m_{\rm e}c^2}\, \left(\frac{3e}{2\pi m_{\rm e}c}\right)^{(p - 1)/2} \nu^{-(p - 1)/2}K_{\rm e},
\end{eqnarray}
$D$ is the distance to the source, and
\begin{equation}
{\cal D} = \frac{1}{\Gamma(1 - {\boldsymbol \beta} {\bf n})}
\end{equation}
is the Doppler factor (${\boldsymbol \beta} = {\bf v}/c$, where ${\bf v}$ is the hydrodynamic velocity and ${\bf n}$ is a unit vector in the observer’s direction). Next, the volume element (which
must already correspond to the laboratory reference
frame in this relation) is written as \mbox{${\rm d}^{3}r = {{\rm d}S \, {\rm d}l}$}, where
${\rm d}S$ is a surface element perpendicular to the line of
sight and ${\rm d}l$ is a length element along the line of sight. Finally, since ${\rm d}S/D^2$ is an element of the solid angle, for the brightness temperature $T_{\rm br} = I_{\nu}c^2/(2 k_{\rm B}\nu^2)$ we ultimately obtain
\begin{eqnarray}
T_{\rm br} = R(p)  \frac{e^3}{m_{\rm e}k_{\rm B}} \, \left(\frac{e}{m_{\rm
e}c}\right)^{(p-1)/2} \nu^{-(p+3)/2} \times \nonumber \\
\times \int {\cal D}^{2 + (p -1)/2} h^{(p +
1)/2} n_{\gamma}({\bf r}) \, \times \nonumber \\
\times \gamma_{0}^{p - 1}({\bf r})(\sin{\hat \chi})^{(p + 1)/2} {\rm d}l,
\end{eqnarray}
where now $n_{\gamma}({\bf r})$ is the particle number density in the
laboratory frame, $k_{\rm B}$ is the Boltzmann constant,
\begin{eqnarray}
R(p) = \frac{3^{p/2}}{8(2\pi)^{(p-3)/2}} \frac{(p - 1)(p + 7/3)}{(p + 1)} \times \nonumber \\
\times \Gamma\left(\frac{3p - 1}{12}\right)\Gamma\left(\frac{3p + 7}{12}\right),
\end{eqnarray}
and the integral is taken along the line of sight.

Next, let us again express the number density of radiating
particles $n_{\gamma}$ via the Goldreich-Julian number density $n_{\rm GJ}$ (2):
\begin{equation}
n_{\gamma} = \lambda_{\gamma} n_{\rm GJ},
\end{equation}
where the constant $\lambda_{\gamma}$,
\begin{equation}
\lambda_{\gamma} \ll \lambda,
\end{equation}
is the multiplicity of radiating particles, while at $\theta \ll 1$, when $z \approx r$,
\begin{eqnarray}
n_{\rm GJ} = \frac{ B_{\rm L}}{2\pi e}
\frac{R_{\rm L}}{r^2}\cos\theta
\approx 1.1\cdot 10^{-5}\,\mbox{cm}^{-3} \times \nonumber \\
\times \left(\frac{h_0}{100\,\mbox{G}}\right)
\left(\frac{R_{\rm L}}{10 \, r_{\rm g}}\right)^{-1}
\left(\frac{M_{\rm bh}}{10^9 M_{\odot}}\right)^{-1}
\left(\frac{z}{R_{\rm L}}\right)^{-2}
\label{nGJconical}
\end{eqnarray}
for electromagnetic fields with a conical shape of the
magnetic surfaces~(\ref{Conical_field_1})--(\ref{Conical_field_3}) and
\begin{eqnarray}
n_{\rm GJ} = \frac{B_{\rm L}}{\sqrt{5} \pi e r} \approx
1.0\cdot 10^{-5}\,\mbox{cm}^{-3} \times \nonumber \\
\times \left(\frac{h_0}{100\,\mbox{G}}\right)
\left(\frac{R_{\rm L}}{10 \, r_{\rm g}}\right)^{-1}
\left(\frac{M_{\rm bh}}{10^9 M_{\odot}}\right)^{-1}
\left(\frac{z}{R_{\rm L}}\right)^{-1}   
\label{nGJparabolic} 
\end{eqnarray}
for electromagnetic fields with a parabolic shape of the
magnetic surfaces~(\ref{parabolBpol}), (\ref{parabolE}), and (\ref{parabolB}) at $X/R_{\rm L} < 0.5$
and constant $\Omega_{\rm F}= c/R_{\rm L}$.

As a result, we obtain
\begin{eqnarray}
T_{\rm br} = \lambda_{\gamma} \frac{m_{\rm e}c^2}{k_{\rm B}} \frac{R(p)}{2
\pi}  \left(\frac{eh_{0}}{m_{\rm e}c}\right)^{(p+3)/2} \nu^{-(p+3)/2} \times \nonumber \\
\times \int {\cal D}^{2 + (p -1)/2} \left(\frac{h}{h_{0}}\right)^{p} \left(\frac{z}{R_{\rm
L}}\right)^{-b} (\sin{\hat \chi})^{(p + 1)/2} 
\frac{{\rm d}l}{R_{\rm L}}.
\end{eqnarray}
Here, $h_{0}$ is the value of $h$ on the light cylinder, while
$b = 2$ and $b = 1$ for conical and parabolic flows, respectively (these values correspond to the laws of decrease in the poloidal magnetic field for these models). In addition, we use the typical mass of a central black hole $M_{\rm bh} = 10^9 M_{\odot}$ with the corresponding Schwarzschild radius $r_{\rm g}=3\cdot 10^{14}\,\mbox{cm}$.

Let us estimate explicitly the optical depth $\tau \sim \mu_{l} l$ for synchrotron self-absorption by relativistic
particles with the power-law energy distributions~(\ref{ne})
and~(\ref{Ke_exp}). Here, $\mu_{l}$ is the synchrotron self-absorption
coefficient and $l \sim \theta_{\rm jet}z$ is the characteristic length.
Using the standard expression for $\mu_{l}$ (see, e.g.,
Zheleznyakov 1997), for the parameters considered
above we obtain
\begin{eqnarray}
\tau = 0.2 \, 
\left(\frac{\lambda_{\gamma}}{10^9}\right)  
\left(\frac{\nu}{15\,\mbox{GHz}}\right)^{-(p+4)/2}  \times \nonumber \\
\times \left(\frac{h_{0}}{100\,\mbox{G}}\right)^{(p+4)/2} 
\left(\frac{z}{10 \, R_{\rm L}}\right)^{1-b(p+4)/2} 
\left(\frac{\theta_{\rm jet}}{0.1} \right)
\mbox{.} 
\label{kappa_s}
\end{eqnarray}
Here, we again used the expression $n_{\gamma} = \lambda_{\gamma}n_{\rm GJ}$ to
determine the number density of radiating particles
and set the lower cutoff limit of their spectrum $\gamma_0 \sim 1$. As we see, for typical $p=2$, $h_{0} = 100 \,\mbox{G}$, and
observation frequency $\nu = 15\,\mbox{GHz}$ the optical depth
for synchrotron self-absorption is much smaller than
unity already at distances $z = 10 \, R_{\rm L}$ from the central engine. Thus, we may neglect the influence
of synchrotron self-absorption on the spectrum and
polarization of the observed radiation of relativistic
electrons on parsec scales.

Regarding the estimate of the Faraday rotation,
it depends significantly on the composition of the
outflowing plasma. In this paper we assume that
the inner paraxial parts of parsec jets are composed
of an electron–positron plasma. This is suggested by both theoretical considerations (Mo{\'s}cibrodzka 
et al. 2011) and observational constraints (Nokhrina
et al. 2015). The presence of positrons reduces dramatically the Faraday rotation. However, it is difficult
to give its quantitative theoretical estimate, since the
exact composition of the jets is unknown. Therefore,
we will again make reference to the rotation measure
observations that point to the absence of noticeable
depolarization in the centimeter wavelength range
even in the innermost central regions of parsec jets.

\section*{Discussion of results}

\begin{figure*}[!ht]
	\begin{minipage}{0.5\linewidth}
		\center{\includegraphics[width=1\linewidth]{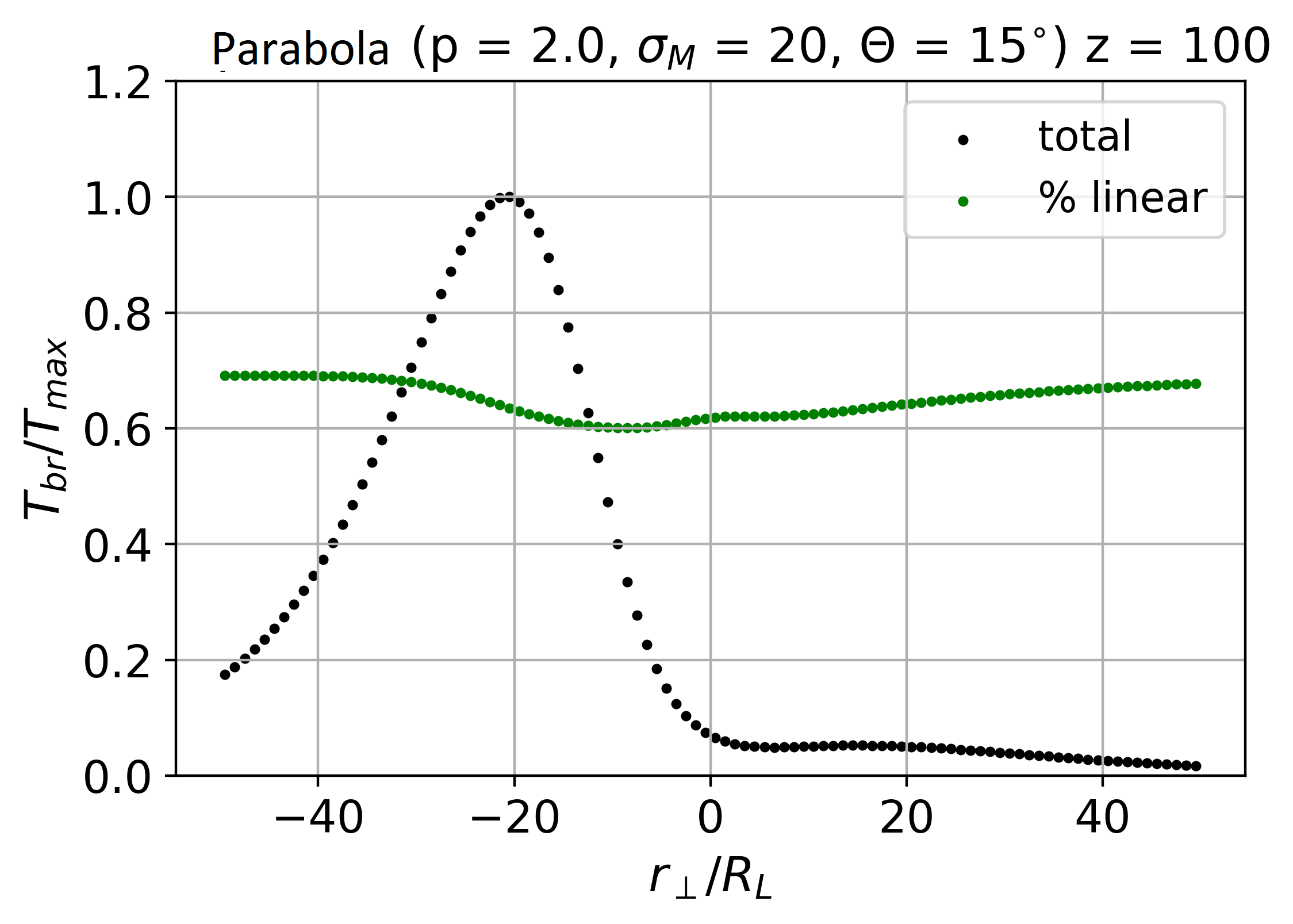} }
	\end{minipage}
	\hfill
	\begin{minipage}{0.5\linewidth}
		\center{\includegraphics[width=1\linewidth]{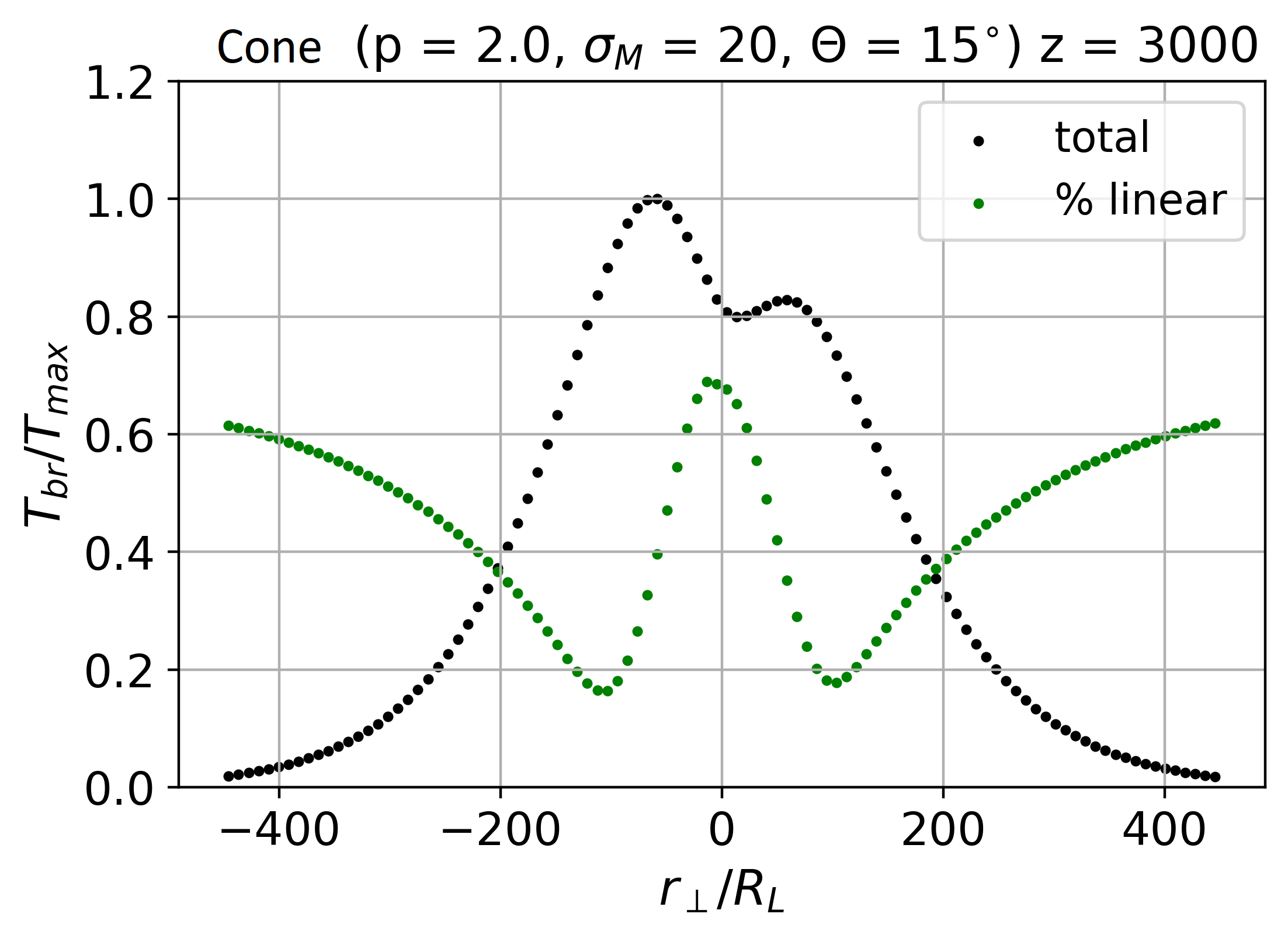} }
	\end{minipage}
 	\caption{Typical brightness temperature profiles for parabolic and conical flows. 
  The degree of linear polarization at 15 GHz is also shown.
  }
\label{fg3}	
\end{figure*}

Figure~\ref{fg3} shows typical brightness temperature
profiles for parabolic (small distances) and conical
(great distances) jets at 15 GHz, for which both 
self-absorption and Faraday rotation may be neglected.
The double-humped profile arises due to the increase
in the hydrodynamic particle energy with distance
from the jet axis, which leads to an increase in the
Doppler factor ${\cal D}$. The asymmetry is attributable to
the jet rotation, owing to which the Doppler factor is
larger in the part of the jet where the rotation is in
the direction to the observer. As we see, this effect
is particularly pronounced at small distances. Such
an asymmetry in the brightness temperature, along
with a W shape of the degree of linear polarization at
great distances $z$, is indeed observed (Butuzova and
Pushkarev 2023).

However, a detailed comparison of the theoretical
and observed transverse brightness temperature profiles is beyond the scope of this paper; a separate paper
will be devoted to such a comparison. Here, having
made sure that our results are valid, we consider
only the dependence of the brightness temperature
$T_{\rm br}$ on the distance $z$ to the central engine. As is
usually done, we use the maximum brightness temperature in cross section (i.e., along the so-called
ridge line).

\begin{table}
\begin{center}
\caption{Exponents $a$ in the relation $T_{\rm br} \propto z^{-a}$ before and
after the break for a conical flow for two values of the angle
$\Theta$ between the jet axis and the direction to the observer.
The jet half-angle is $\theta_{\rm jet} = 6^{\circ}$}
\vspace{0.3cm}
\begin{tabular}{c|cccc}
\hline
$\sigma_{\rm M}$  & 10 & 20 & 30 & 40 \\
\hline
$\Theta = 2^{\circ}$ & & & & \\
$p = 2.0$ & 2.7$-$3.0 & 3.0$-$3.0 & 3.3$-$3.0 & 3.6$-$3.0  \\
$p = 2.5$ & 3.2$-$3.5 & 3.7$-$3.5 & 4.0$-$3.5 & 4.3$-$3.5  \\
$p = 3.0$ & 3.6$-$3.8 & 4.1$-$3.8 & 4.4$-$3.8 & 4.7$-$3.8 \\

\hline
$\Theta = 20^{\circ}$ & & & &  \\
$p = 2.0$ & 4.2$-$3.0 & 5.6$-$3.0 & 6.2$-$3.0 & 6.5$-$3.0  \\
$p = 2.5$ & 4.9$-$3.6 & 6.5$-$3.5 & 6.7$-$3.6 & 6.9$-$3.6  \\
$p = 3.0$ & 6.0$-$4.0 & 7.4$-$4.2 & 7.7$-$4.3 & 7.8$-$4.5  \\

\hline
\end{tabular}
\label{tab1}
\end{center}
\end{table}

\begin{table}
\begin{center}
\caption{Exponents $a$ in the relation $T_{\rm br} \propto z^{-a}$ before
and after the break for a parabolic flow for two values of
the angle $\Theta$ between the jet axis and the direction to the
observer.}
\vspace{0.3cm}
\begin{tabular}{c|cccc}
\hline
$\sigma_{\rm M}$  & 10 & 20 & 30 & 40  \\
\hline
$\Theta = 2^{\circ}$ & & & & \\
$p = 2.0$ & 1.4$-$2.6 & 1.5$-$3.0 & 1.5$-$3.0 & 1.5$-$3.0 \\
$p = 2.5$ & 1.7$-$3.1 & 1.8$-$3.1 & 1.8$-$3.1 & 1.9$-$3.1 \\
$p = 3.0$ & 2.2$-$3.4 & 2.0$-$3.2 & 2.0$-$3.2 & 2.1$-$3.2 \\
\hline
$\Theta = 20^{\circ}$ & & & & \\
$p = 2.0$ & 3.1$-$2.1 & 2.9$-$2.0 & 2.9$-$2.0 & 2.9$-$2.0 \\
$p = 2.5$ & 3.4$-$2.3 & 3.1$-$2.3 & 3.1$-$2.3 & 3.1$-$2.3  \\
$p = 3.0$ & 3.6$-$2.7 & 3.3$-$2.7 & 3.3$-$2.7 & 3.3$-$2.7  \\
\hline
\end{tabular}
\label{tab2}
\end{center}
\end{table}

\begin{figure*}[!ht]
	\begin{minipage}{0.5\linewidth}
		\center{\includegraphics[width=1\linewidth]{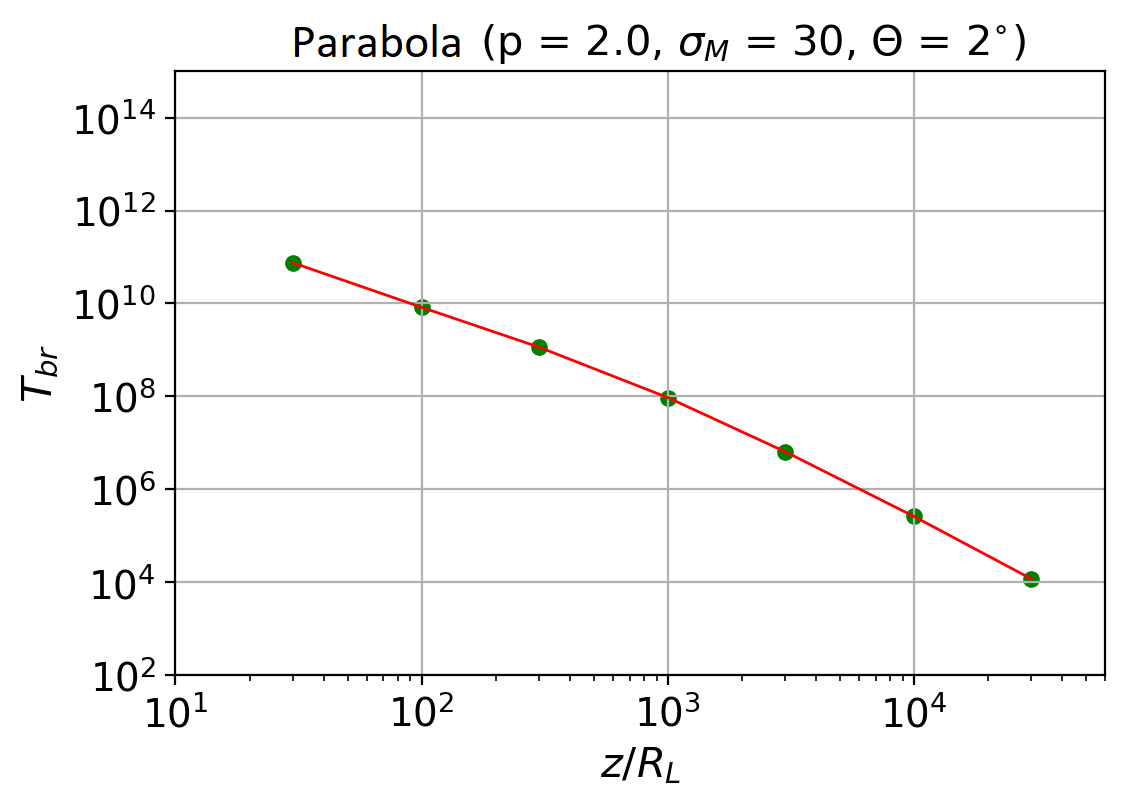} }
	\end{minipage}
	\hfill
	\begin{minipage}{0.5\linewidth}
		\center{\includegraphics[width=1\linewidth]{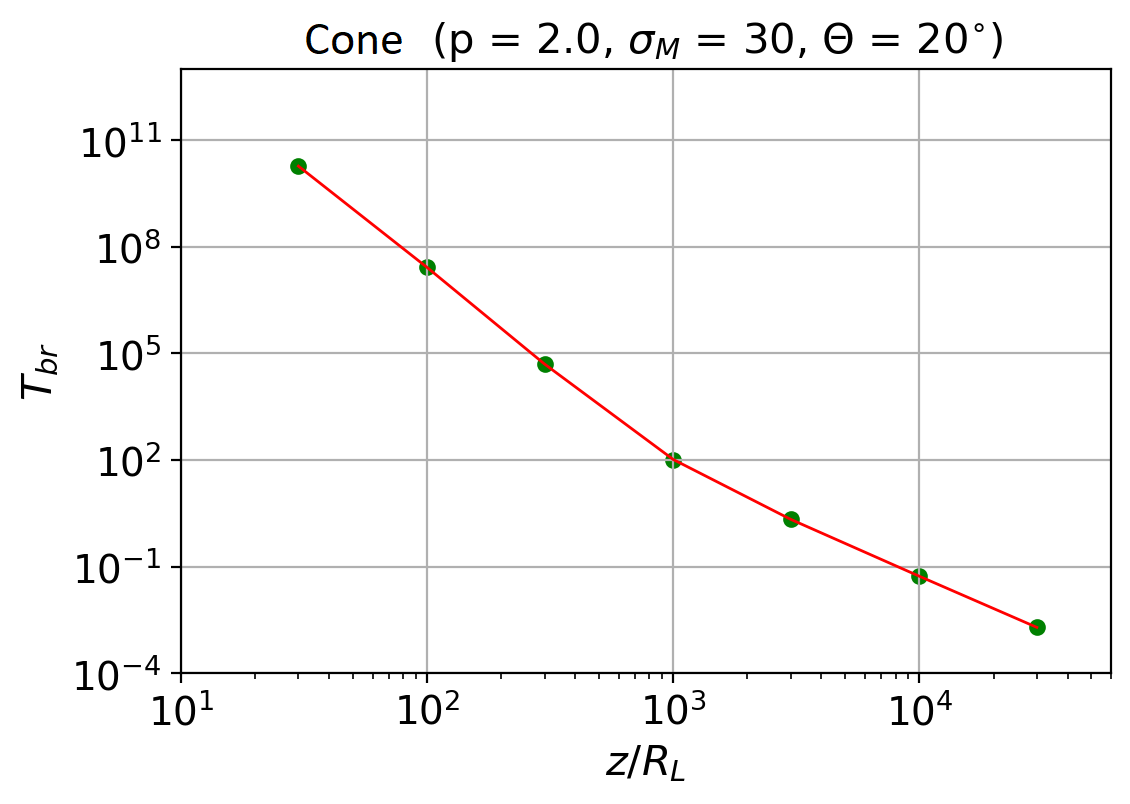} }
	\end{minipage}
 	\caption{Maximum of the brightness temperature $T_{\rm br}$ (in degrees)
         versus distance $z$ to the central engine. In both cases, $\lambda_{\gamma}/\lambda = 0.001$. 
         For typical black hole masses $\sim 10^9 M_{\odot}$ and $R_{\rm L} \sim 10 r_{\rm g}$ the position of the break corresponds to observed distances $\sim$ 1 pc.
         }
\label{fg4}	
\end{figure*}

For typical parameters of relativistic jets $\sigma_{\rm M} = 20$
and observation angle $\Theta = 20^{\circ}$ Fig.~\ref{fg4} shows two
examples of the dependence of the brightness temperature $T_{\rm br}$ 
(in degrees) on the distance to the central engine. For typical black 
hole masses $\sim 10^{9}M_{\odot}$ and commonly adopted 
$R_{\rm L} \sim 10 \, r_{\rm g}$ the position of the break exactly 
corresponds to distances $\sim 1$ pc. A full summary of our results 
for the exponents $a$ in the dependence $T_{\rm br} \propto z^{-a}$ 
before and after the breakfor two values of the angle $\Theta$ between 
the jet axis and the direction to the observer is given in Tables~\ref{tab1}
and \ref{tab2}. As we see, these dependences do have a break in the 
saturation region even if the jet geometry does not change in the 
saturation region. The value of the break itself depends on both spectral 
index $p$ and magnetization parameter $\sigma_{\rm M}$. The break can
be directed both upward and downward. A fortiori
there will be a break when passing from a parabolic
flow to a conical one. The wide spread of parameters
presented in Tables~\ref{tab1} and \ref{tab2} may well explain the
breaks observed in the distance dependence of the
brightness temperature in jets.

Yet another important point that should be emphasized here is 
that to obtain the brightness temperatures corresponding to 
observations, we set $\lambda_{\gamma} = 10^{9}$--$10^{10}$. 
In other words, to explain the observations, it is sufficient 
that the energy density of the nonthermal particles be lower
than the rest energy density of the cold (background) plasma 
by two or three orders of magnitude. Thus, the model considered 
here receives yet another confirmation.

\section*{Conclusions}

We showed that the conservation of the first adiabatic invariant 
(i.e., the relationship between the magnetic field in the comoving 
reference frame $h$ and the energy of radiating particles) naturally 
leads to a change in the dependence of the brightness temperature 
$T_{\rm br}$ on the distance $z$ to the central engine, since
there are different asymptotic behavior of $h(z)$ before and after
the saturation region. This effect was demonstrated
for both parabolic and conical relativistic jet structures.
Both the presence of a break and the characteristic behavior
of the linear polarization qualitatively
reproduce the observational data.

Note once again that the goal of this paper was to
show only the fundamental possibility of a change in
brightness temperature along the jet axis related to
the change in the spectrum of radiating particles due
to the conservation of the first adiabatic invariant. 
In contrast, a detailed comparison with observations
will become possible only after the refinement of
a number of circumstances that were ignored in the
above analysis.

This primarily concerns allowance for the transverse inhomogeneity of jets 
(see, e.g., Nokhrina et al. 2015), which must result in the noticeable fall
off of the particle number density and the poloidal magnetic field  toward 
the flow periphery. Accordingly, the dependence of $\sigma_{\rm M}$ (and, 
hence, the parameter $\varepsilon$) on the distance to the axis should also 
be taken into account. This can lead to a significant change
in brightness temperature. Yet another factor that
can also affect quantitatively the results obtained is
related to the synchrotron losses. In the Appendix we
discussed the losses of particles whose energies correspond to observed frequencies $\sim 10$ GHz. It is clear,
however, that the energy losses will be significant for
sufficiently high energies of radiating particles, which
will lead to a change in the high-energy part of the
spectrum. In turn, if the first adiabatic invariant
is conserved, at large $z$ this region of the spectrum
will already correspond to the observed frequencies.
It is not inconceivable that this effect can explain
$a > 4$ observed in a number of sources. Finally,
in this paper we did not discuss the low-frequency
self-absorption that can also affect significantly the
observed brightness temperature of jets.

\section*{Acknowledgments}

We are grateful to E. Kravchenko, M. Lisakov, A. Lobanov,
and I. Pashchenko for the useful discussion. We also
thank the two anonymous referees whose critical remarks
contributed to a refinement of the argumentation of our
model and allowed a number of inaccuracies to be removed.
This study was supported by the Russian Foundation for
Basic Research (project no. 20-02-00469).

\section*{Appendix}

Above we implicitly assumed that the synchrotron
losses of radiating particles could be neglected. Here we
will show that this assumption does hold. For this purpose,
let us define the quantity $\epsilon$ in the plasma rest frame,
\begin{equation}
\epsilon = \frac{\left({\rm d}\gamma^{\prime}/{\rm  d}t^{\prime}\right)_{\rm syn}}{\left({\rm d}\gamma^{\prime}/{\rm  d}t^{\prime}\right)_{\rm inv}},
\end{equation}
where the derivative
\begin{equation}
\left(\frac{{\rm d}\gamma^{\prime}}{{\rm d}t^{\prime}}\right)_{\rm syn} = \frac{2}{3} \, \frac{e^4h^2}{m_{\rm e}^3c^5}\gamma^2
\end{equation}
corresponds to the synchrotron losses at $\chi = 90^{\circ}$, while for
any power-law dependences of $h$ on $z$
\begin{equation}
\left(\frac{{\rm d}\gamma^{\prime}}{{\rm d}t^{\prime}}\right)_{\rm inv} \approx \frac{\Gamma c \gamma^{\prime}}{z}.
\end{equation}
Here, for simplicity, we substituted ${\rm d}t = \Gamma {\rm d}t^{\prime}$
 and $c \, {\rm d}t = {\rm d}z$. As a result, we obtain
\begin{equation}
\epsilon \approx \frac{\omega_{h}^2 r_{\rm e}z \gamma^{\prime}}{\Gamma c^2},
\end{equation}
where $\omega_{h} = e h/m_{\rm e}c$. To estimate this quantity, it should
be remembered that we are interested in the observations
at a fixed frequency $\nu \approx \omega_{h} (\gamma^{\prime})^2 \Gamma$ and, therefore,
\begin{equation}
\epsilon \propto z \, h^{3/2}(z) \Gamma^{-3/2}(z).
\end{equation}
As a result, we obtain in the region before the break $x < 1/\sqrt(2\varepsilon)$ (and neglecting the dependence on $\Gamma$)
\begin{eqnarray}
\epsilon_{\rm conus} & \propto & z^{-2}, 
\label{var1}\\
\epsilon_{\rm parabolic} & \propto & z^{-1/2}.
\label{var2}
\end{eqnarray}
As we see, in both cases the parameter $\epsilon$ decreases with
increasing $z$. On the other hand, for $\nu = 15$ GHz
\begin{equation}
\epsilon(R_{\rm L}) \sim 1 \left(\frac{B_{\rm L}}{10^2 \, G}\right)^{3/2} \left(\frac{R_{\rm L}}{10^{15} \, {\rm cm}}\right).
\end{equation}

\end{document}